\documentclass[aps,showpacs,preprintnumbers,amsmath,amssymb]{revtex4}
\UseRawInputEncoding
\usepackage{dcolumn}

\usepackage{float}
\usepackage{graphics,epsfig}
\usepackage{epstopdf}
\usepackage{graphicx}
\usepackage{multirow}
\usepackage{dcolumn}
\usepackage{bm}
\usepackage{gensymb}

\begin{document}
\baselineskip=0.8 cm
\title{\bf Kerr black hole shadows from axion-photon coupling }
\author{
Songbai Chen$^{1,2}$\footnote{Corresponding author: csb3752@hunnu.edu.cn},
Jiliang Jing$^{1,2}$ \footnote{jljing@hunnu.edu.cn}}
\affiliation{ $ ^1$ Department of Physics, Institute of Interdisciplinary Studies, Key Laboratory of Low Dimensional Quantum Structures
    and Quantum Control of Ministry of Education, Synergetic Innovation Center for Quantum Effects and Applications, Hunan
    Normal University,  Changsha, Hunan 410081, People's Republic of China
    \\
    $ ^2$Center for Gravitation and Cosmology, College of Physical Science and Technology, Yangzhou University, Yangzhou 225009, People's Republic of China}

\begin{abstract}
\baselineskip=0.6 cm
\begin{center}
{\bf Abstract}
\end{center}
We have investigated the motion for photons in the  Kerr black hole spacetime under the axion-photon coupling. The birefringence phenomena arising from the axion-photon coupling can be negligible in the weak coupling approximation because the leading-order contributions to the equations of motion come from the square term of the coupling parameter.
We find that the coupling parameter makes the size of shadows slightly increase for arbitrary spin parameter. For the rapid rotating black hole case with a larger coupling, we find that there exist a ``pedicel"-like structure appeared in the left of the ``D"-type like shadows.
Comparing the shadow size of the Kerr black hole with the shadow size of the Sgr A* and M87* black holes,  we make constraints on the parameter space for such a theoretical model of the axion-photon coupling.

\end{abstract}

 \maketitle
\newpage
\section{Introduction}

The images of the supermassive black holes M87* \cite{fbhs1,fbhs6,fbhs12,fbhs13} and Sgr A* \cite{fbhs1222,fbhs17} open a new window to test gravity in strong field regimes. Meanwhile, they also provide a powerful way to probe  electromagnetic interactions, matter distributions and accretion processes near black holes
\cite{kerr1,kerr2,bhs0,bhs1,bhs2,bhs3,bhsp1,swo11,Vagnozzi, min22,shadow2022}. With the circularity
and size of its first image, the rotational nature of the supermassive object M87* was tested \cite{com1} and the Kerr black hole hypothesis was further examined \cite{testkerr}. The image features of black holes with extra hairs have been extensively studied \cite{scalar1,scalar2,scalar3,scalar4,non2,non4}, which could provide a way to check no-hair theorem in the strong gravity region. Moreover, the supermassive black hole images have been applied to probe the Lorentz symmetry violation \cite{non3,non5,non8}, hunt extra dimensions \cite{non6,non6a1}, test Loop quantum gravity \cite{non7,non7a1,non7a2} and examine other alternative theories \cite{non9,non10}. The  black hole images  have also been studied in the Euler-Heisenberg and Bronnikov non-linear electrodynamics models coupled to general relativity \cite{non1}.
The main component in the black hole's image is the shadow, which is caused by light rays that fall into
an outer event horizon. In general, the black hole shadow depends on the parameters of black hole backgrounds, the propagations of
light rays and the positions of observers.
Recent studies also show that the some interactions between electromagnetic and
gravitational fields leads to birefringence of photons in spacetimes, which result in double shadows for a single black hole \cite{qed1,qed2,qed3,weyl6,weyl7,weyl6s}.
These interesting
features have triggered the further study of black hole shadows under interactions between
electromagnetic and other fields.  A phenomenological coupling between a photon and a generic vector field is also introduced to study black hole shadows \cite{epb}, which shows that the black hole
shadow in edge-on view  has different appearances for
different frequencies of the observed lights.

Dark matter is widely believed to be the dominant gravitationally attractive component in the Universe although its nature is still unclear.
One of the most interesting candidates for dark matter is axion, which is hypothetical pseudoscalar particle  initially
introduced by Peccei and Quinn \cite{cpqus} to solve the strong charge-conjugation and parity problem in quantum
chromodynamics. Interestingly, axion is also furthermore generically predicted in string theory \cite{Arvanitaki,Witten,Marsh}. One of the important properties of axions is that they
interact with photons through the coupling, which leads to an interesting conversion from photons into axions
and vice versa in the presence of a magnetic field \cite{Maiani,Raffelt}. Such axion-photon conversion
is also regarded as basic principle \cite{Sikivie,Irastorza} to experimentally detect
Solar axions \cite{Anastassopoulos,Armengaud,Banerjee} and axion dark matter \cite{Asztalos}. Moreover, to
account for the recent detections of high-energy gamma ray
photons from extragalactic sources, some suggestions \cite{Mirizzi,Meyer} based on the axion-photon conversion are introduced because the conversion can prevent the high-energy photons from being annihilated through electron-positron pair production in their propagations. Recent investigations \cite{Yanagida,Mirizzi1,Tashiro} show that the resonant axion-photon conversion inside the cluster's magnetic field could  distort the black-body spectrum of the cosmic microwave background.

As a coupling between matter and electromagnetic fields, the axion-photon coupling also results in the photon birefringence \cite{Harari} as photon
crosses over axion matter. The birefringence effects of the polarization photons have been used to analyze
the axion dark matter distribution near M87* black hole \cite{tomoch} and in the protoplanetary disk around a young star \cite{Fujita}.
The axion-induced birefringence effect gives arise to the unique polarimetric structure \cite{Alexander} and the electric vector position angle oscillation of linearly polarized photons \cite{Yifan}. Such oscillation has been applied to constrain  region of the axion mass and axion-photon coupling parameter space together with the observation data from Event Horizon Telescope \cite{Yifan}. Moreover, the conversion of  photons into axions leads to a dimming of the photon ring  around the black hole shadow \cite{Kimihiro}. The photon scattering from the background magnetic field with axions is also found to generate a significant circular polarization around the horizon
of supermassive black hole \cite{Soroush} and in blazars \cite{Run}. The polarization-dependent
bending that a ray of light experiences by traveling through an axion cloud is studied in
the background of a Kerr black hole \cite{Plascencia}. All above literature is focused  on analyzing effects of the axion-photon coupling on the brightness and polarization patterns of surrounding emissions region in black hole images. However, it is still an open issue how the axion-photon coupling affect black hole shadows.
In this paper, we start from the modified Maxwell equation and make use of the geometric optics approximation to obtain the equation of motion for the photon coupling to an axion field. Then, we probe the effects of axion-photon coupling on the shadow of a rotating black hole.

The paper is organized as follows: In Sec.II,  we firstly present equation of motion for photons interacted with axion-like particles in the Kerr black hole spacetime and obtain two kind of solutions of polarized photon motions. In Sec.III, we present numerically Kerr black hole shadows under the axion-photon coupling and probe its effects on the shadows. Finally, we end the paper with a summary.

\section{Equation of motion for the photons coupling axion-like particles in a Kerr black hole spacetime}

We firstly present the equations of motions for photons interacted with axion-like particles
in a Kerr black hole spacetime by the geometric optics approximation
\cite{weyl0,Daniels,Caip,Cho1,Lorenci}. In the curved spacetime,
the action contained the coupling between photon and axion-like particle
 can be expressed as \cite{cpqus,Maiani,Raffelt,Sikivie,Irastorza,Harari}
\begin{eqnarray}
S=\int d^4x \sqrt{-g}\bigg[\frac{R}{16\pi
G}+\frac{1}{2}\partial_{\mu}\psi\partial^{\mu}\psi-\frac{1}{4}\bigg(F_{\mu\nu}F^{\mu\nu}-\alpha \psi
F_{\mu\nu}\tilde{F}^{\mu\nu}\bigg)\bigg],\label{acts}
\end{eqnarray}
where $\psi$ denotes the axion-like field and $\tilde{F}^{\mu\nu}=\epsilon^{\mu\nu\rho\sigma}F_{\rho\sigma}/2$ is the dual of the electromagnetic field strength tensor. $\epsilon^{\mu\nu\rho\sigma}$ is the Levi-Civita tensor and $\alpha$ denotes the axion-photon coupling
constant with dimensions of inverse energy. Generally, the axion-like field $\psi$ is dynamical. Here, we assume it as a function of a radial and polar angle coordinate $\psi=\psi(r, \theta)$  for the sake of simplicity. Therefore, the coupling  between photon and axion-like particle modifies Maxwell equation as
\begin{eqnarray}
\nabla_{\mu}F^{\mu\nu}-2\alpha \psi_{\mu}
\epsilon^{\mu\nu\rho\sigma}F_{\rho\sigma}=0,\label{WE}
\end{eqnarray}
where $\psi_{\mu}\equiv\partial_{\mu}\psi$.
Using the geometric optics approximation \cite{weyl0,Daniels,Caip,Cho1,Lorenci}, we can get equation of motions for coupled photons from the modified Maxwell equation (\ref{WE}). In this approximation, the wavelengths of photons are assumed to be much smaller than a typical curvature
scale, but be larger than the electron Compton wavelength. Then, the electromagnetic tensor $F_{\mu\nu}$ can be rewritten as a simpler form
\begin{eqnarray}
F_{\mu\nu}=f_{\mu\nu}e^{i\theta},\label{ef1}
\end{eqnarray}
with a slowly varying amplitude $f_{\mu\nu}$ and a rapidly varying phase $\theta$. Comparing with  the wave vector  $k_{\mu}=\partial_{\mu}\theta$, one can find that the derivative term $f_{\mu\nu;\lambda}$ can be neglected since it is not dominated in this case. Making use of the Bianchi identity
\begin{eqnarray}
D_{\lambda} F_{\mu\nu}+D_{\mu} F_{\nu\lambda}+D_{\nu} F_{\lambda\mu}=0,
\end{eqnarray}
it is easy to obtain that the form of the amplitude $f_{\mu\nu}$  must be
\begin{eqnarray}
f_{\mu\nu}=k_{\mu}a_{\nu}-k_{\nu}a_{\mu}.\label{ef2}
\end{eqnarray}
Here the polarization vector $a_{\mu}$ is orthogonal to the wave vector $k_{\mu}$, i.e.,
$k_{\mu}a^{\mu}=0$.
Inserting Eqs.(\ref{ef1}) and (\ref{ef2}) into Eq. (\ref{WE}), we can obtain the modified equation of motions for photons under the axion-photon coupling
\begin{eqnarray}
k_{\mu}k^{\mu}a^{\nu}-i4\alpha \psi_{\mu}\epsilon^{\mu\nu\rho\sigma}k_{\sigma}a_{\rho}=0,\label{WE2}
\end{eqnarray}
which means that the coupling  will change the propagation of photons in background spacetimes.

With the standard Boyer-Lindquist coordinates,   the  metric of  a Kerr black hole can be expressed as
\begin{eqnarray}
ds^2&=&-\rho^2\frac{\Delta}{\Sigma^2}dt^2+\frac{\rho^2}{\Delta}dr^2+\rho^2
d\theta^2+\frac{\Sigma^2}{\rho^2}\sin^2{\theta}(d\phi-\omega dt)^2,\label{m1}
\end{eqnarray}
with
\begin{eqnarray}
\omega&=&\frac{2aMr}{\Sigma^2},
\;\;\;\;\;\;\;\;\;\;\;\;\;\;\;\;\;\;\;\;\;\;\;\;\;\;\;\;\;\rho^2=r^2+a^2\cos^2\theta,
\nonumber\\
\Delta&=&r^2-2Mr+a^2,\;\;\;\;\;\;\;\;\;\;\;\;\;\;\;\;\Sigma^2=(r^2+a^2)^2-a^2\sin^2\theta \Delta.
\end{eqnarray}
Here the parameters $M$ and $a$ denote the mass and the spin parameter of the black hole, respectively.
For the Kerr black hole spacetime, it is convenient to
build a local set of orthonormal frames by introducing vierbein fields $e^a_{\mu}$ obeyed the condition
\begin{eqnarray}
g_{\mu\nu}=\eta_{ab}e^a_{\mu}e^b_{\nu},
\end{eqnarray}
where $\eta_{ab}$ denotes the Minkowski metric. The forms of the vierbein fields $e^a_{\mu}$ and their inverse $e_a^{\mu}$  can be respectively expressed as
\begin{eqnarray}
e^a_{\mu}=\left(\begin{array}{cccc}
\rho\frac{\sqrt{\Delta}}{\Sigma}&0&0&-\frac{\omega\Sigma}{\rho}\sin\theta\\
0&\frac{\rho}{\sqrt{\Delta}}&0&0\\
0&0&\rho&0\\
0&0&0&\frac{\Sigma}{\rho}\sin\theta
\end{array}\right),
\end{eqnarray}
and
\begin{eqnarray}
e_a^{\mu}=\left(\begin{array}{cccc}
\frac{\Sigma}{\rho\sqrt{\Delta}}&0&0&0\\
0&\frac{\sqrt{\Delta}}{\rho}&0&0\\
0&0&\frac{1}{\rho}&0\\
\frac{\omega\Sigma}{\rho\sqrt{\Delta}}&0&0&\frac{\rho}{\Sigma\sin\theta}
\end{array}\right).
\end{eqnarray}
Making use of the relationship $a^{\mu}k_{\mu}=0$, one can find that the equation of motion of the
photon coupling with axion can be simplified as a set of equations for three independent
polarisation components $a^{r}$, $a^{\theta}$, and $a^{\phi}$,
\begin{eqnarray}
\bigg(\begin{array}{ccc}
K_{11}&K_{12}&K_{13}\\
K_{21}&K_{22}&
K_{23}\\
K_{31}&K_{32}&K_{33}
\end{array}\bigg)
\bigg(\begin{array}{c}
a^{r}\\
a^{\theta}
\\
a^{\phi}
\end{array}\bigg)=0.\label{Kk1}
\end{eqnarray}
The coefficients $K_{ij}$ are very complicated and here we do not list them. The necessary and sufficient condition for Eq.(\ref{Kk1}) to have non-zero solutions
 is that the determinant of its coefficient matrix is zero. Solving the equation $|K|=0$, we
  obtain two non-zero physical solutions,
 \begin{eqnarray}
g_{\mu\nu}\dot{x^{\mu}}\dot{x^{\nu}}\pm 4\alpha\sqrt{(\psi_{\mu}\dot{x^{\mu}})^2+4\alpha^2(\psi_{\mu}\psi^{\mu})^2}+8\alpha^2\psi_{\mu}\psi^{\mu}=0.  \label{solution2s}
\end{eqnarray}
This means that the coupling between photon and axion-like scalar field could give arise to birefringence phenomena, even if in the non-rotating case. The light cone condition (\ref{solution2}) implies that the coupling photons  propagate along non-geodesic paths in the Kerr spacetime.
Considered that the coupling parameter $\alpha$ is small for  physically justification, the left side of the equation (\ref{solution2s}) can be expanded as an Taylor series near $\alpha=0$. Neglecting its higher terms $\mathcal{O}(\alpha^3)$, we find that the equation (\ref{solution2s}) can be further approximated  as
\begin{eqnarray}
g_{\mu\nu}\dot{x^{\mu}}\dot{x^{\nu}}\pm 4\alpha\psi_{\mu}\dot{x^{\mu}}+8\alpha^2\psi_{\mu}\psi^{\mu}=0.  \label{solution2}
\end{eqnarray}
Thus, the motion of the coupling photon can be actually determined by a Lagrange function
\begin{eqnarray}
\mathcal{L}=\frac{1}{2}g_{\mu\nu}\dot{x^{\mu}}\dot{x^{\nu}}\pm 2\alpha\psi_{\mu}\dot{x^{\mu}}+4\alpha^2\psi_{\mu}\psi^{\mu}. \label{lageri1}
\end{eqnarray}

\begin{figure}
\includegraphics[width=6cm]{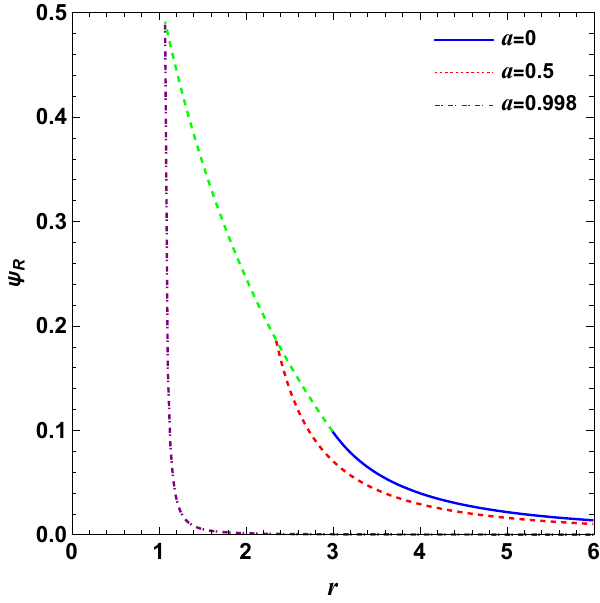}
\caption{ Change of the radial part $\psi_R$ of the scalar field in Eq.(\ref{scfield}) with the coordinate $r$ in the region outside the marginally unstable circular orbit of photons for different spin parameter $a$. The green dashed line corresponds to the curve $\psi_R-r_{ph}$, where $r_{ph}$ is the marginally unstable circular radius of photon. }\label{fig0}
\end{figure}

From the previous discussion, under the geometric optics approximation \cite{weyl0,Daniels,Caip,Cho1,Lorenci}, we have $F_{\mu\nu}=f_{\mu\nu}e^{i\theta}=(k_{\mu}a_{\nu}-k_{\nu}a_{\mu})e^{i\theta}$ and then the coupling term $\alpha F_{\mu\nu}\tilde{F}^{\mu\nu}$ in the Klein-Gordon equation becomes
\begin{eqnarray}
\alpha F_{\mu\nu}\tilde{F}^{\mu\nu}=\frac{\alpha }{2}\epsilon^{\mu\nu\rho\sigma}F_{\mu\nu} F_{\rho\sigma}=2\alpha \epsilon^{\mu\nu\rho\sigma}k_{\mu}a_{\nu}k_{\rho}a_{\sigma}e^{2i\theta}=0,
\end{eqnarray}
which means that under geometric optics approximation, the coupling photons do not change the distribution of the scalar field in the spacetime. Here, we assume that the scalar field possesses the same symmetry as the Kerr spacetime and has a form $\psi=\psi(r,\theta)$,  then the corresponding
 Klein-Gordon equation becomes
\begin{eqnarray}
\frac{1}{\sqrt{-g}}\frac{d}{d r}\bigg(\sqrt{-g}g^{rr} \frac{d \psi(r,\theta)}{d r}\bigg)+\frac{1}{\sqrt{-g}}\frac{d}{d \theta}\bigg(\sqrt{-g}g^{\theta\theta} \frac{d \psi(r,\theta)}{d \theta}\bigg)=0. \label{coscalar}
\end{eqnarray}
Assuming the scalar field has a form $\psi(r,\theta)=\mathcal{R}(r)\Theta(\theta)$ in the Kerr spacetime, we have
\begin{eqnarray}
\frac{1}{\mathcal{R}(r)}\frac{d}{d r}\bigg(\Delta \frac{d \mathcal{R}(r)}{d r}\bigg)=-\frac{1}{\Theta(\theta)\sin\theta}\frac{d}{d \theta}\bigg(\sin\theta \frac{d \Theta(\theta)}{d \theta}\bigg)=l(l+1).\label{coscalar1}
\end{eqnarray}
It is easy to find that the general solution of Eq. (\ref{coscalar1}) is
\begin{eqnarray}
\mathcal{R}(r)=C_1P_l(\frac{r-M}{\sqrt{M^2-a^2}})+C_2Q_l(\frac{r-M}{\sqrt{M^2-a^2}}),\;\;\;\;\;\;\;\;\Theta(\theta)=C_3P_l(\cos\theta)+C_4Q_l(\cos\theta).
\end{eqnarray}
Where $P_l(x)$ and $Q_l(x)$ are the Legendre functions of the first and second kinds, respectively. $C_i$ ($i=1,2,3,4)$ are the corresponding integration constants. For the sake of simplicity, we set $l=1$, $C_1=\frac{\pi i}{2} C_2$, $C_2=C_3=C$ and $C_4=0$, where $C$ is assumed to be positive without loss of generality. This special solution of scalar field is not divergent at the spatial infinity  and has a form
\begin{eqnarray}
\psi(r,\theta)=C\bigg[\frac{r-M}{2\sqrt{M^2-a^2}}\ln \frac{r-r_-}{r-r_+}-1\bigg]\cos\theta,\label{scfield}
\end{eqnarray}
where $r_{\pm}=M\pm\sqrt{M^2-a^2}$. The radial part $\psi_R\equiv\frac{r-M}{2\sqrt{M^2-a^2}}\ln \frac{r-r_-}{r-r_+}-1$ decreases with the coordinate $r$. The boundaries of black hole shadows depend on the marginally unstable circular orbits of photons. From Fig.\ref{fig0}, we find that
the absolute value $|\psi_R|<1$ in the region outside the marginally unstable circular orbit of photons even if in the rapidly rotation case.
Thus, in the propagation of lights reaching the spatial infinity, the scalar field meets the condition $|\psi|<C$.
With the form of scalar field in Eq. (\ref{scfield}),  we can obtain its derivative with respect to coordinates $x^{\mu}$
\begin{eqnarray}
\psi{_{\mu}}=C\bigg[0,\;\; \bigg(\frac{1}{2\sqrt{M^2-a^2}}\ln \frac{r-r_-}{r-r_+}-\frac{r-M}{\Delta}\bigg)\cos\theta, \;\;-\bigg(\frac{r-M}{2\sqrt{M^2-a^2}}\ln \frac{r-r_-}{r-r_+}-1\bigg)\sin\theta, \;\;0\bigg].\label{coscalar2}
\end{eqnarray}

Inserting Eq.(\ref{coscalar2}) into Eq.(\ref{lageri1}), we find that the Lagrange function $\mathcal{L}$  is independent of the coordinates $t$ and $\phi$, the photon's energy $E_0$ and its $z$-component of the angular momentum $L_{z0}$ are two conserved quantities as in the case without the coupling. With these two conserved quantities and the Lagrange-Euler equation,
we can obtain the equation of motion of the coupling photon
\begin{eqnarray}
\dot{t}&=&\frac{g_{\phi\phi}E_0+g_{t\phi}L_{z0}}{g_{t\phi}^2-g_{tt}g_{\phi\phi}},\quad\quad\quad\quad\quad
\dot{\phi}=\frac{g_{t\phi}E_0+g_{tt}L_{z0}}{g_{tt}g_{\phi\phi}-g_{t\phi}^2},\label{u1}\\
\ddot{r}&=&
\frac{1}{2g_{rr}}\bigg[g_{tt,r}\dot{t}^2-g_{rr,r}\dot{r}^2-2g_{rr,\theta}\dot{r}\dot{\theta}+g_{\theta\theta,r}\dot{\theta}^2+g_{\phi\phi,r}\dot{\phi}^2 +2g_{t\phi,r}\dot{t}\dot{\phi}+8\alpha^2\frac{\partial }{\partial r}\bigg(\psi{_{\mu}}\psi^{\mu}\bigg) \bigg],
\label{uu2}\\
\ddot{\theta}&=&\frac{1}{2g_{\theta\theta}}\bigg[g_{tt,\theta}\dot{t}^2+g_{rr,\theta}\dot{r}^2-
2g_{\theta\theta,r}\dot{r}\dot{\theta}
-g_{\theta\theta,\theta}\dot{\theta}^2+g_{\phi\phi,\theta}\dot{\phi}^2+2g_{t\phi,\theta}\dot{t}\dot{\phi}+8\alpha^2\frac{\partial }{\partial \theta}\bigg(\psi{_{\mu}}\psi^{\mu}\bigg) \bigg].
\label{uu3}
\end{eqnarray}
Here we note that the term $\pm 2\alpha\psi_{\mu}\dot{x^{\mu}}$ has no contribution to the equations of motion of the coupling photons since the force arising from this term is $\pm 2\alpha g^{\mu\nu}(\psi_{\rho,\mu}-\psi_{\mu,\rho})\dot{x}^{\rho}$, which is zero because the scalar field $\psi$ is a continuous function and its second-order partial derivative is independent of the sequence of derivation. The leading-order contributions to the equations of motion  (Eqs.(\ref{uu2}) and (\ref{uu3})) come from the corrected terms contained the factor $\alpha^2$, which implies that the birefringence phenomena may be negligible in the small $\alpha$ approximation. In the next section, we will study the effects of the axion-photon coupling on Kerr black hole shadows.

\section{Shadow of Kerr black hole under the axion-photon coupling}

 To obtain the shadow of Kerr black hole under the axion-photon coupling, we must adopt the ``backward ray-tracing" method as in \cite{scalar1,scalar2,scalar3,scalar4,binary,sha18,my,swo7,swo8,swo9,swo10} because the motion equations (\ref{u1})-(\ref{uu3}) can not be variable-separable.
 With this method, the position of each pixel in the final image can be got by solving numerically the nonlinear differential equations (\ref{u1})-(\ref{uu3}) by
 assuming that light rays evolve from the observer backward in time. The black hole shadow in observer's sky is determined only by the pixels related to the light rays falling down into black hole.
For the Kerr spacetime (\ref{m1}), the transformation between the local basis of observer
$\{e_{\hat{t}},e_{\hat{r}},e_{\hat{\theta}},e_{\hat{\phi}}\}$  and the coordinate basis of the background spacetime $\{ \partial_t,\partial_r,\partial_{\theta},\partial_{\phi} \}$ can be expressed as
\begin{eqnarray}
\label{zbbh}
e_{\hat{\mu}}=e^{\nu}_{\hat{\mu}} \partial_{\nu},
\end{eqnarray}
and the transformation matrix  $e^{\nu}_{\hat{\mu}}$ satisfies $g_{\mu\nu}e^{\mu}_{\hat{\alpha}}e^{\nu}_{\hat{\beta}}
=\eta_{\hat{\alpha}\hat{\beta}}$, where $\eta_{\hat{\alpha}\hat{\beta}}$ is the usual Minkowski metric. For the Kerr spacetime (\ref{m1}), one can conveniently choose the transformation matrix $e^{\nu}_{\hat{\mu}}$ as
\begin{eqnarray}
\label{zbbh1}
e^{\nu}_{\hat{\mu}}=\left(\begin{array}{cccc}
\zeta&0&0&\gamma\\
0&A^r&0&0\\
0&0&A^{\theta}&0\\
0&0&0&A^{\phi}
\end{array}\right),
\end{eqnarray}
where $\zeta$, $\gamma$, $A^r$, $A^{\theta}$,and $A^{\phi}$ are real coefficients. The decomposition (\ref{zbbh1}) is actually connected with a reference frame with zero axial angular momentum in relation to spatial infinity \cite{sw,swo,astro,chaotic,binary,sha18,my,swo7,swo8,swo9,swo10}.
According to the Minkowski normalization
\begin{eqnarray}
e_{\hat{\mu}}e^{\hat{\nu}}=\delta_{\hat{\mu}}^{\hat{\nu}},
\end{eqnarray}
we obtain the coefficients in the matrix $e^{\nu}_{\hat{\mu}}$ as
\begin{eqnarray}
\label{xs}
&&A^r=\frac{1}{\sqrt{g_{rr}}},\;\;\;\;\;\;\;\;\;\;\;\;\;\;
A^{\theta}=\frac{1}{\sqrt{g_{\theta\theta}}},\;\;\;\;\;\;\;\;\;\;\;\;\;\;
A^{\phi}=\frac{1}{\sqrt{g_{\phi\phi}}},\nonumber\\
&&\zeta=\sqrt{\frac{g_{\phi\phi}}{g^2_{t\phi}-g_{tt}g_{\phi\phi}}},\;\;\;\;\;\;\;\;\;\;\;\;\;\;\;\;\;\;\;\;
\gamma=-\frac{g_{t\phi}}{g_{\phi\phi}}\sqrt{\frac{g_{\phi\phi}}{g^2_{t\phi}-g_{tt}g_{\phi\phi}}},
\end{eqnarray}
With Eq.(\ref{zbbh}), we further obtain the locally measured four-momentum $p^{\hat{\mu}}$ of photons in the Kerr spacetime (\ref{m1})  as
\begin{eqnarray}
\label{kmbh}
p^{\hat{t}}&=&\zeta E_0-\gamma L_{z0},\;\;\;\;\;\;\;\;p^{\hat{\phi}}=\frac{1}{\sqrt{g_{\phi\phi}}}p_{\phi},
\;\;\;\;\;\;\;\;
p^{\hat{\theta}}=\frac{1}{\sqrt{g_{\theta\theta}}}p_{\theta},
\;\;\;\;\;\;\;\;\;\;
p^{\hat{r}}=\frac{1}{\sqrt{g_{rr}}}p_{r}.
\end{eqnarray}
Thus, the corresponding celestial coordinates for pixels corresponding to light rays can be expressed as
\begin{eqnarray}
\label{xd1}
x=-r_{obs}\frac{p^{\hat{\phi}}}{p^{\hat{r}}}
=-r_{obs}\frac{\sqrt{g_{rr}}(g_{t\phi}\dot{t}+g_{\phi\phi}\dot{\phi})}{\sqrt{g_{\phi\phi}}(g_{rr}\dot{r}+2\alpha \psi_{r})}, \quad\quad\quad
y=r_{obs}\frac{p^{\hat{\theta}}}{p^{\hat{r}}}=
r_{obs}\frac{\sqrt{g_{rr}}(g_{\theta\theta}\dot{\theta}+2\alpha \psi_{\theta})}{\sqrt{g_{\theta\theta}}(g_{rr}\dot{r}+2\alpha \psi_{r})}.
\end{eqnarray}
where $r_{obs}$ and $\theta_{obs}$ are respectively the radial coordinate and polar angle of observer.
\begin{figure}
\centering
\includegraphics[width=12cm]{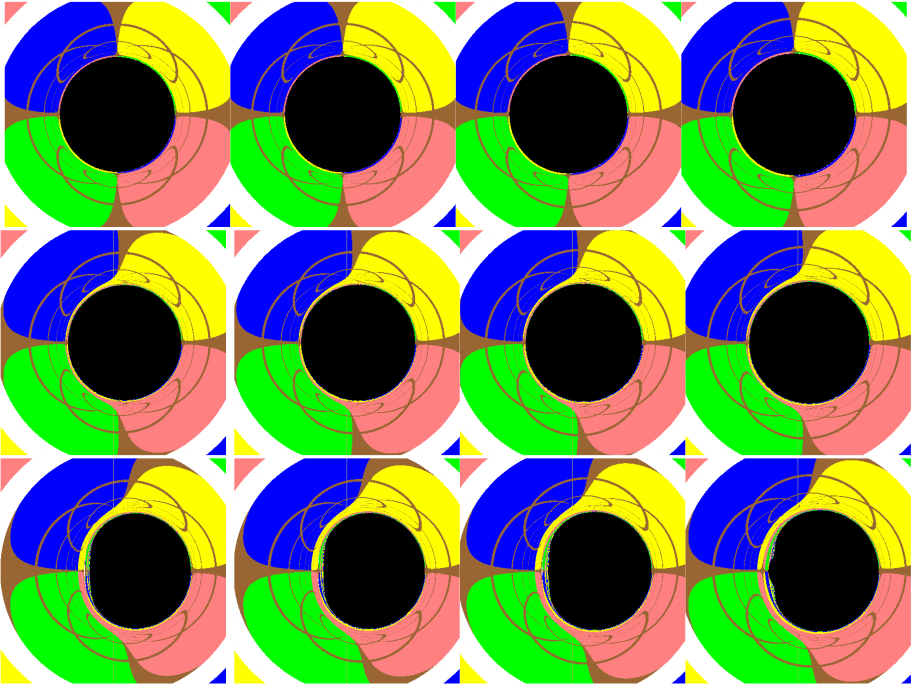}
\caption{ Shadows of a Kerr black hole casted by the extraordinary rays with different coupling parameter $\tilde{\alpha}$ and spin parameter $a$ for the observer in the equatorial plane ($\theta_{obs}=\pi/2$). The top, middle and bottom rows respectively denote the cases with $a=0$, $0.5$, and $0.998$. In each row, the figures  from left to right correspond to $\tilde{\alpha}=0$, $0.1$, $0.2$  and $0.3$, respectively. Here we set the mass parameter $M=1$ and $r_{obs}=50M$.  }
\label{fig1}
\end{figure}

\begin{figure}
\centering
\includegraphics[width=12cm]{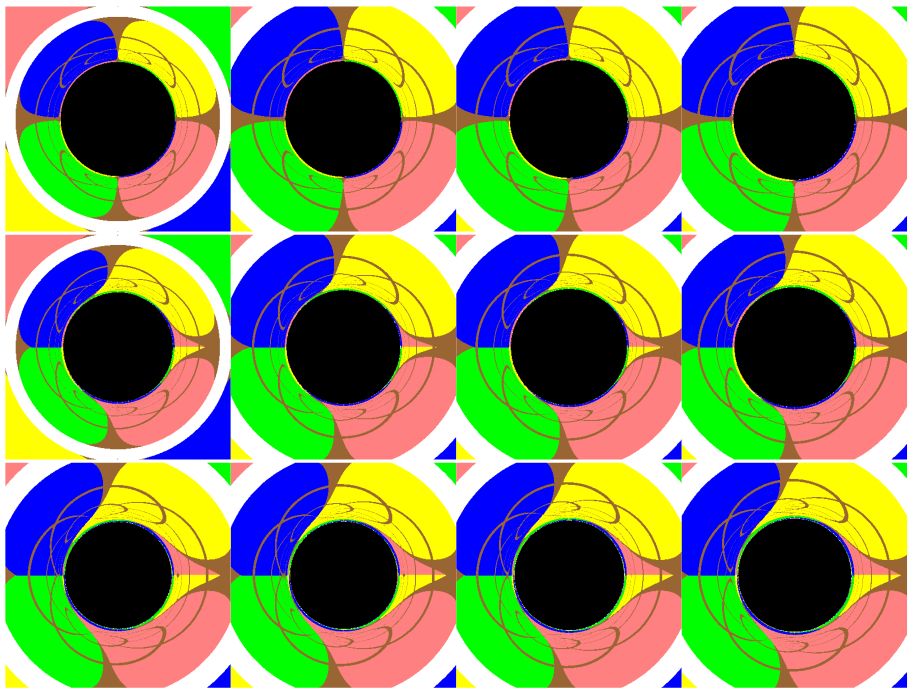}
\caption{  Shadows of a Kerr black hole casted by the extraordinary rays with different coupling parameter $\tilde{\alpha}$ and spin parameter $a$ for the observer  along the rotation axis of black hole ($\theta_{obs}=0^{\circ}$). The top, middle and bottom rows respectively denote the cases with $a=0$, $0.5$, and $0.998$. In each row, the figures  from left to right correspond to $\tilde{\alpha}=0$, $0.1$, $0.2$  and $0.3$, respectively. Here we set the mass parameter $M=1$ and $r_{obs}=50M$. }
\label{fig2}
\end{figure}

Figs.\ref{fig1} and \ref{fig2} present Kerr black hole shadows under the axion-photon coupling for different spin parameters.
To demonstrate the deformation near the black hole arising from strong gravitational lensing, we divide the total celestial sphere into four quadrants painted in different colors (green, blue,
red, and yellow) \cite{sw,swo,astro,chaotic,binary,sha18,my,swo7,swo8,swo9,swo10} and draw the grids of longitude and latitude lines marked with adjacent brown
lines separated by $10^\circ$.  Actually, the points with different colors in each panel are images of the source points lied in the four different quadrants,  which entirely demonstrate the deformation  near the black hole originating from strong gravitational lensing.
The black parts denote black hole shadows and the white rings provide a direct demonstration of Einstein rings.

Fig.\ref{fig1} presents the Kerr black hole shadows observed in the equatorial plane. It shows that the coupling parameter $\tilde{\alpha}=\alpha C$ results in a larger size of shadows for arbitrary spin parameter $a$. Moreover, for the rapid rotating black hole, its shadow has a usual ``D" type shape as the coupling parameter $\tilde{\alpha}$ is smaller. With the increasing of $\tilde{\alpha}$, we find that there exist a ``pedicel"-like structure
gradually appeared in the left of the shadow, which is similar to those in shadows of a disformal Kerr black hole in quadratic degenerate
higher-order scalar-tensor (DHOST) theory \cite{fen10}.
As the observer stands along the direction of the rotation axis of black hole ($\theta_{obs}=0^{\circ}$), the black hole shadows in Fig.\ref{fig2}  have a shape of a circular disk. With the increase of $\tilde{\alpha}$, the size of shadows slightly increases. This is also consistent with the effects of $\tilde{\alpha}$ on the shadow observed in the equatorial plane.

\begin{figure}
\includegraphics[width=5cm]{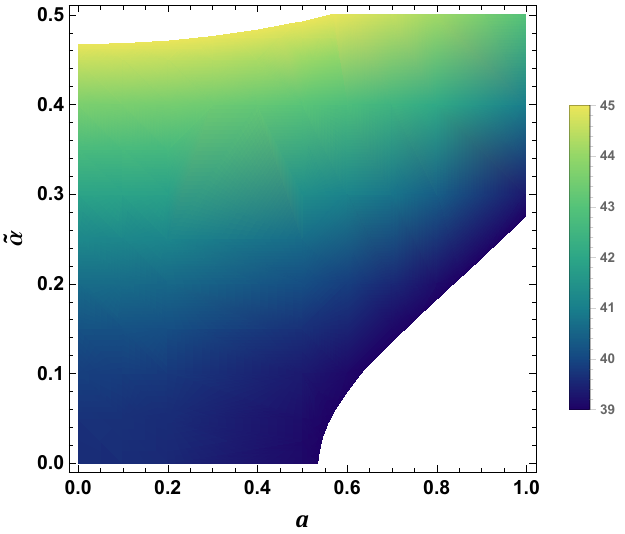}\includegraphics[width=5cm]{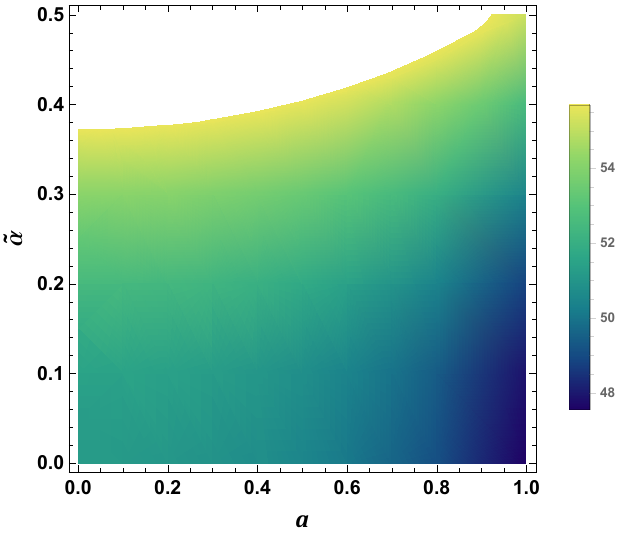}\includegraphics[width=5cm]{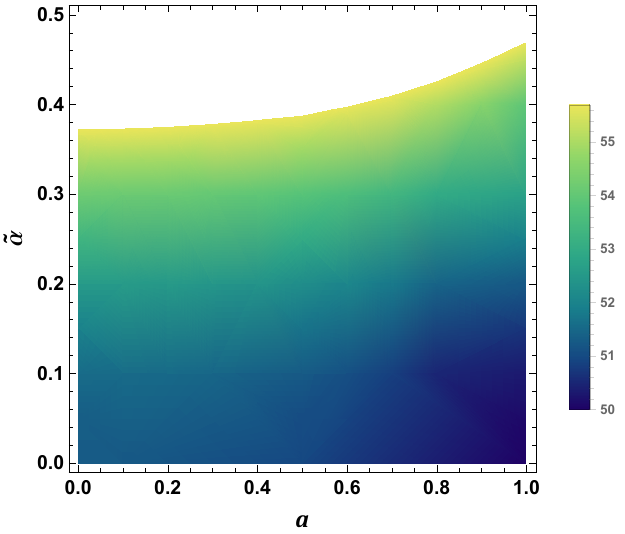}
\caption{ The permissible parameter region (colored) $a-\tilde{\alpha}$ for the model with axion-photon interaction around black holes using the M87* and Sgr A* shadow images. The left panel is obtained by using the M87* image, where the boundary is estimated by $ 39 \mu as \leq \theta_d \leq 45 \mu as$. The middle panel and the right one are drawn by using the Sgr A*  image with the observed inclination angle $\theta_0=0^{\circ}$ and $\theta_0=50^{\circ}$, respectively. The boundaries in the last two panels are estimated by $ 41.7 \mu as \leq \theta_d \leq 55.7 \mu as$. }
\label{fig2tt}
\end{figure}

Let us now to make constraints on the parameter space of the model with axion-photon interaction around black holes using the M87* and Sgr A* images released by the Event Horizon Telescope (EHT) Collaboration. In left panel in Fig. \ref{fig2tt}, the permissible parameter region (colored) $a-\tilde{\alpha}$ for the model is obtained by using the M87* shadow images with the black hole mass $M=6.5\times10^{9}M_{\odot}$, the observer distance $D_{O}=16.8Mpc$ and the inclination angle $\theta_0=17^{\circ}$. The boundaries of the permissible parameter region are estimated by the angular diameter $ 39 \mu as \leq \theta_d \leq 45 \mu as$ \cite{fbhs1}.
Since the size of shadows increases with the coupling parameter $\tilde{\alpha}$ and decreases with the black hole spin parameter $a$, the constraints from the M87*
show that the smaller coupling parameter is allowable and the larger one is excluded in the case where the black hole is slowly spinning, and the situation is exactly the opposite in the rapidly spinning black hole case. The middle and right panels show the constraints on the parameter space $a-\tilde{\alpha}$ of the model from the Sgr A* shadows with the angular diameter $ 41.7 \mu as \leq \theta_d \leq 55.7 \mu as$ \cite{fbhs1222}. Here
we use the mass $M =4.0\times10^{6}M_{\odot}$ and the observer distance $D_{O}=8.0kpc$ for the black hole Sgr A*. Unlike the case M87*, there is no clear consensus on the inclination angle $\theta_0$ for the Sgr A* and the current images of the Sgr A* from EHT rule out that the black
hole is viewed at high inclination ($\theta_0 \geq 50^{\circ}$ ) \cite{fbhs1222}. Therefore, in Fig.\ref{fig2tt}, we set the observed inclination angle $\theta_0=0^{\circ}$ in the middle panel and $\theta_0=50^{\circ}$ in the right one to get the largest allowable parameter region. The results show that  the allowable range of $\tilde{\alpha}$ increases with the black hole spin.
In addition, in the rapidly spinning case, the larger inclination angle gives a tighter constraint
on the parameter $\tilde{\alpha}$. From Fig.\ref{fig2tt}, it is easy to
find that the tightest upper limit of the coupling constant is $\tilde{\alpha}\lesssim 0.5$. The CERN Axion Solar Telescope (CAST) provides a limit on the axion-photon coupling strength $\alpha <0.66\times10^{-10}{\rm GeV}$$^{-1}$ from globular cluster stars \cite{Anastassopoulos}.  A upper limit on the axion-photon coupling $\alpha <5.3\times10^{-12}{\rm GeV}$$^{-1}$  is recently obtained from the supernova SN1987A  \cite{addtest1}. Analyzing the observational data from the Perseus Galaxy Cluster collected with the MAGIC telescopes for NGC 1275,  one can find that the limit for axion-like particles is $\alpha <3\times10^{-12}{\rm GeV}$$^{-1}$ \cite{addtest2}.
These constraints from observations are performed for the axion-photon coupling constant with dimensions of inverse energy. In this work, the coupling constant  $\tilde{\alpha}=\alpha C$ is actually a dimensionless axion photon coupling constant like $c=2\pi \alpha f_a$, where $f_a$ is the Peccei-Quinn symmetry breaking scale. With data from  EHT polarimetric measurements of M87*, one obtains the constraint on the
dimensionless coupling constant $c<0.1$ \cite{tomoch,Yifan}. Thus, our constraint is a little weaker than that obtained in \cite{tomoch,Yifan}. These results could help to understand deeply the axion field and their corresponding interactions with electromagnetic field.

\section{summary}

We have investigated the motion for photons in the  Kerr
black hole spacetime under the axion-photon coupling.
Although the axion-photon coupling yields birefringence,  the birefringence phenomena can be negligible in the small $\alpha$ approximation because the leading-order contributions to the equations of motion come from the corrected terms contained the factor $\alpha^2$.
We also probe the effects of the coupling on the black hole shadows. It shows that the coupling parameter $\alpha$  makes the size of shadows slightly increase for arbitrary spin parameter $a$. Moreover, for the rapid rotating black hole, its shadow observed in the equatorial plane has a usual ``D" type shape as the coupling parameter $\tilde{\alpha}$ is smaller. With the increasing of $\tilde{\alpha}$, we find that there exist a ``pedicel"-like structure
gradually appeared in the left of the shadow. Finally, we compare the shadow size of the Kerr black hole with those of the Sgr A* and M87* black holes with different coupling values of $\tilde{\alpha}$ and find that there is room for such a theoretical model of the axion-photon coupling.
Moreover, in the limit of geometric optics, it is found  \cite{Dominik}
that the birefringence due to axionic fields is found to be achromatic, but the redshift of light and distance estimates based on propagating light rays, and the shear and magnification due to gravitational lensing are not affected by the interaction of light with an axionic field.
These results could help to deeply understand  the axion field and their corresponding interactions with electromagnetic field.

\section{\bf Acknowledgments}
We thank the referee for the important and insightful
comments which are very helpful in the improvement of
this work. This work was  supported by the National Natural Science
Foundation of China under Grant No.12275078, 11875026, 12035005, and 2020YFC2201400, and the innovative research group of Hunan Province under Grant No. 2024JJ1006.

\end{document}